\pdfoutput=1
\documentclass[%preprint,
natbib,10pt]{TOOLS/sigplanconf}
\usepackage{amsmath}
\usepackage{graphicx}
\usepackage{subfig}
\usepackage{url}
\usepackage{verbatim} 
\usepackage{listings}
\lstnewenvironment{codex}
    {\lstset{}%
      \csname lst@SetFirstLabel\endcsname}
    {\csname lst@SaveFirstLabel\endcsname}
\lstnewenvironment{code}
    {\lstset{}%
      \csname lst@SetFirstLabel\endcsname}
    {\csname lst@SaveFirstLabel\endcsname}
\newtheorem{prop}{Proposition}
\newtheorem{df}{Definition}

\newcommand{\BI}[0]{\begin{itemize}}
\newcommand{\EI}[0]{\end{itemize}}

\newcommand{\BE}[0]{\begin{enumerate}}
\newcommand{\EE}[0]{\end{enumerate}}

\newcommand{\BX}[0]{\begin{codex}}
\newcommand{\EX}[0]{\end{codex}}
\newcommand{\BV}[0]{\begin{verbatim}}
\newcommand{\EV}[0]{\end{verbatim}}
    
\def \bscale1 {0.50}
\def \bscale {0.25}

\newcommand{\FIG}[3]{
\begin{figure}[htbp]
\centering
\scalebox{\bscale1}{\includegraphics*[bb=0pt 0pt 800pt 800pt]{./figs/#3.png}}
\caption{#2}
\label{#1}
\end{figure}
}

\newcommand{\VFIGS}[6]{
\begin{figure}[htbp]
  \begin{center}
    {\scalebox{\bscale}{\includegraphics*[bb=0pt 0pt 800pt 800pt]{./figs/#5.png}}}
    {\scalebox{\bscale}{\includegraphics*[bb=0pt 0pt 800pt 800pt]{./figs/#6.png}}}
  \caption{#2: {\em #3} and {\em #4}}
  \label{#1}
  \end{center}
\end{figure}
}

\begin{document}

\conferenceinfo{WXYZ '08}{date, City.} 
\copyrightyear{2008} 
\copyrightdata{[to be supplied]} 

\titlebanner{banner above paper title}        % These are ignored unless
\preprintfooter{short description of paper}   % 'preprint' option specified.

\title{
    A Functional Hitchhiker's Guide to Hereditarily Finite Sets, 
    Ackermann Encodings and Pairing Functions
}
\subtitle{{\em -- unpublished draft --}}
           
\authorinfo{Paul Tarau}
   {Department of Computer Science and Engineering\\
   University of North Texas}
   {\em tarau@cs.unt.edu}

\maketitle

\date{}

\begin{abstract}
The paper is organized as a self-contained literate Haskell program that
implements elements of an executable finite set theory with focus on
combinatorial generation and arithmetic encodings. The code, 
tested under GHC 6.6.1, is available at
\url{http://logic.csci.unt.edu/tarau/research/2008/fSET.zip}.

We introduce ranking and unranking functions generalizing Ackermann's
encoding to the universe of Hereditarily Finite Sets 
with Urelements. Then we build a lazy enumerator for 
Hereditarily Finite Sets with Urelements 
that matches the unranking function provided by
the inverse of Ackermann's encoding and
we describe functors between them
resulting in arithmetic encodings 
for powersets, hypergraphs, ordinals and
choice functions. 
After implementing a digraph representation of 
Hereditarily Finite Sets we define {\em decoration functions} 
that can recover well-founded sets 
from encodings of their associated acyclic digraphs.
We conclude with an encoding of arbitrary digraphs and discuss 
a concept of duality induced by the set membership relation.
%In the process, we uncover the surprising possibility of internally sharing 
%isomorphic objects, independently of their language level types and meanings.

\keywords
{\em 
hereditarily finite sets, 
ranking and unranking functions, 
executable set theory,
arithmetic encodings, 
Haskell data representations,
functional programming and computational mathematics
}
\end{abstract}

\section{Introduction}
While the Universe of Hereditarily Finite Sets is best known 
as a model of the Zermelo-Fraenkel Set theory with the 
Axiom of Infinity  replaced by its 
negation \cite{finitemath,DBLP:journals/jct/MeirMM83}, 
it has been the object of renewed practical interest in 
various fields, from representing structured data in 
databases \cite{DBLP:conf/foiks/LeontjevS00} to reasoning 
with sets and set constraints in a Logic Programming framework 
\cite{dovier00comparing,DBLP:journals/tplp/PiazzaP04,DBLP:conf/cav/DovierPP01}.

The Universe of Hereditarily Finite Sets is built from the 
empty set (or a set of {\em Urelements}) by successively 
applying powerset and set union operations.
A surprising bijection, discovered by Wilhelm Ackermann 
in 1937 \cite{ackencoding,abian78,kaye07} from 
Hereditarily Finite Sets to Natural Numbers, 
was the original trigger for our
work on building in a mathematically elegant 
programming language, a concise and {\em executable} 
hereditarily finite set theory.  
The arbitrary size of the data objects brought in 
the need for arbitrary length integers.  
The focus on potentially infinite enumerations brought in the need 
for lazy evaluation. These have made {\em Haskell} a natural choice.

We will describe our constructs in a subset of {\em Haskell} 
\cite{haskell:98rr,DBLP:journals/jfp/Jones03,DBLP:journals/jfp/Jones03g} 
seen as a concrete syntax for a generic lambda calculus 
based functional language\footnote{As a courtesy to the reader wondering 
about the title, the author confesses being a hitchhiker 
in the world of functional programming, coming from the 
not so distant galaxy of logic programming but still 
confused by recent hitchhiking trips in the exotic worlds 
of logic synthesis, foundations of mathematics, 
natural language processing, 
conversational agents and virtual reality. And not being afraid to
go boldly where \ldots a few others have already been before.}.

We will only make the assumptions 
that non-strict functions (higher order included), with call-by-need 
evaluation and arbitrary length integers are available in the language. 
While our code will conform Haskell's type system, we will do that 
without any type declarations, by ensuring that the types of our 
functions are all inferred. This increases chances that the code 
can be ported, through simple syntax transformations, 
to any programming language that implements our basic assumptions. 

The paper is organized as follows: section \ref{bit} 
introduces the reader to combinatorial generation 
with help of a bitstring example, section
\ref{ack} introduces Ackermann's encoding in the more general case 
when {\em urelements} are present and shows an encoding for hypergraphs 
as a particular case. Section \ref{obv} gives examples of transporting 
common set and natural number operations from one side to the other. 
After discussing some classic pairing functions, section \ref{pairings} 
introduces new pairing/unpairing on natural numbers.
Section \ref{graphs} discusses graph representations 
and {\em decoration functions} on Hereditarily Finite Sets (\ref{graphrep}), 
and provides encodings for directed acyclic graphs (\ref{digraphs}). 
Sections \ref{related} and \ref{concl} discuss related work, future work and
conclusions.

\begin{comment}
\begin{code}
module FSET where
import Data.List
import Data.Bits
import Data.Array
import Data.Graph
import Random
\end{code}
\end{comment}
\section{What's in a Bit?} \label{bit}
Let us observe first that the well known bitstring representation 
of Natural Numbers (see {\tt to\_rbits} and {\tt from\_rbit} in Appendix and
notice the reversed bit order) is a first hint at their genuinely
polymorphic, ``shapeshifting'' nature:
\begin{verbatim}
to_rbits 2008
  [0,0,0,1,1,0,1,1,1,1,1]
from_rbits [0,0,0,1,1,0,1,1,1,1,1]
  2008
\end{verbatim}
\noindent The effect is trivial here - these transformers turn a number into
a list of bits and back. 
One step further, we will now define two one argument {\em functions},
that implement the ``bits'' {\tt o} and {\tt i}:

\begin{code}
o x = 2*x+0
i x = 2*x+1
\end{code}
One can recognize now that 2008 is just the result of composing
``bits'', with a result similar to the result of {\tt from\_rbits}:
\begin{verbatim}
(o.o.o.i.i.o.i.i.i.i.i) 0
  2008
\end{verbatim}
The reader will notice that we have just
``shapeshifted'' to yet another view: 
a number is now a composition of {\em bits, seen as transformers}, 
where each bit does its share by leftshifting the string
one position and then adding its contribution to it.
Note the analogy with Church numerals, which
represent numbers as iterations of function application,
except that here $n$ will only need O($log_2(n)$) of space.

Like with the usual bitstring representation, the dominant digit
is always {\tt 1}, zeros after that have no effect, from where
we can infer that the mapping
between such bitstrings and numbers is not one-to-one. A variant of the
{\em 2-adic bijective} numeral representation fixes this, and
shows one of the simplest bijective mappings from
natural numbers to bitstrings (i.e. the regular language $\{0,1\}^*$):
\begin{code} 
nat2bits = drop_last . to_rbits . succ where
  drop_last bs=genericTake (l-1) bs where
    l=genericLength bs

bits2nat bs = (from_rbits (bs ++ [1]))-1
\end{code}
\begin{verbatim}
nat2bits 42
  [1,1,0,1,0]
bits2nat it
  42
map nat2bits [0..15]
  [[],[0],[1],[0,0],[1,0],[0,1],[1,1],
   [0,0,0],[1,0,0],[0,1,0],[1,1,0],
   [0,0,1],[1,0,1],[0,1,1],[1,1,1],
   [0,0,0,0]]
\end{verbatim}
The last example suggests that we are now able to generate the {\em infinite
stream of all possible bitstrings} simply as as:
\begin{code}
all_bitstrings = map nat2bits [0..]
\end{code}
We will now hitchhike with this {\em design pattern} in our toolbox to a more
interesting universe.

\section{Hereditarily Finite Sets and the Ackermann Encoding} \label{ack}
The Universe of Hereditarily Finite Sets ($HFS$) is built from the empty set (or
a set of {\em Urelements}) by successively applying powerset and set union operations.
Assuming $HFS$ extended with {\em Urelements} (i.e. objects not having any elements), the 
following data type defines a recursive ``rose tree'' for Hereditarily Finite Sets:
\begin{code}
data HFS t = U t | S [HFS t] deriving (Show, Eq)
\end{code}
We will assume that {\em Urelements} are represented as Natural Numbers in {\tt
[0..ulimit-1]}. The constructor {\tt U t} marks {\em Urelements} of type {\tt
t} (usually the arbitrary length Integer 
type in Haskell) and the constructor {\tt S} marks a 
list of recursively built $HFS$ type elements. 
Note that if no elements are used with the {\tt U} constructor, we obtain the 
``pure'' $HFS$ universe by representing the empty set as {\tt S []}.

\subsection{Ackermann's Encoding} 

A surprising bijection, discovered by Wilhelm Ackermann 
in 1937 \cite{ackencoding,abian78,kaye07} 
maps Hereditarily Finite Sets ($HFS$) to Natural Numbers ($Nat$):

\vskip 0.5cm
$f(x)$ = {\tt if} $x=\{\}$ {\tt then} $0$ {\tt else} $\sum_{a \in x}2^{f(a)}$
\vskip 0.5cm

Let us note that Ackermann's encoding can be seen as the recursive application 
of a bijection {\tt set2nat} from finite subsets of $Nat$ to $Nat$, 
that associates to a set of (distinct!) natural numbers a (unique!) natural number.
 
A simple change to Ackermann's mapping, will accomodate
a finite number of Urelements in $[0..u-1]$, as follows:

\vskip 0.5cm
$f_{u}(x)$ = 
{\tt if} $x<u$ 
{\tt then} $x$
{\tt else} $u+\sum_{a\in x}2^{f_{u}(a)}$ 
\vskip 0.5cm

\begin{prop}
For $u \in Nat$ the function $f_{u}$ is a bijection from $Nat$ to
$HFS$ with Urelements in $[0..u-1]$.
\end{prop}

The proof follows from the fact that no sets map to values smaller than
$ulimit$ and that Urelements map into themselves.

With this representation, Ackermann's encoding 
from $HFS$ with Urelements in {\tt [0..ulimit-1]} 
to $Nat$ {\tt hfs2nat\_} becomes:
\begin{code} 
hfs2nat_ _ (U n) = n
hfs2nat_ ulimit (S es) = 
  ulimit + set2nat (map (hfs2nat_ ulimit) es)

set2nat ns = sum (map (2^) ns)
\end{code}
where {\tt set2nat} maps a set of exponents of 2 to the associated sum of
powers of 2.

We can now define
\begin{code}  
hfs2nat = hfs2nat_ urelement_limit

urelement_limit=0
\end{code}
where the constant {\tt urelement\_limit} controls the initial 
segment of $Nat$ to be mapped to {\em Urelements}. Note that to
keep our Haskell code as simple as possible we assume that
{\tt urelement\_limit} is a global parameter that implicitly
fixes the set of Urelements.

To obtain the inverse of the Ackerman encoding, let's first define the 
inverse {\tt nat2set} of the bijection {\tt set2nat}. It decomposes a 
natural number into a list of exponents of 2 (seen as bit positions 
equaling 1 in its bitstring representation, in increasing order).
\begin{code}
nat2set n = nat2right_exps n 0 where
  nat2right_exps 0 _ = []
  nat2right_exps n e = add_rexp (n `mod` 2) e 
    (nat2right_exps (n `div` 2) (e+1)) where
      add_rexp 0 _ es = es
      add_rexp 1 e es = (e:es)
\end{code}
\begin{verbatim}
nat2set 42
  [1,3,5]
set2nat [1,3,5]
  42
nat2set 2008
  [3,4,6,7,8,9,10]
set2nat [3,4,6,7,8,9,10]
  2008
\end{verbatim}
\noindent The inverse of the (bijective) Ackermann encoding (generalized to
work with urelements in {\tt [0..ulimit-1]}) is defined as follows:
\begin{code}
nat2hfs_ ulimit n | n<ulimit = U n 
nat2hfs_ ulimit n = 
  S (map (nat2hfs_ ulimit) (nat2set (n-ulimit)))
\end{code}
We can now define
\begin{code}  
nat2hfs = nat2hfs_ urelement_limit
\end{code}
where the constant {\tt urelement\_limit} controls the initial 
segment of $Nat$ to be mapped to {\em Urelements}.

As both {\tt nat2hfs} and {\tt hfs2nat} are obtained through 
recursive compositions of {\tt nat2set} and {\tt set2nat}, respectively, 
one can generalize the encoding mechanism by replacing these building 
blocks with other bijections with similar properties.

One can try out {\tt nat2hfs} and its inverse {\tt hfs2nat} and 
print out a $HFS$ with the {\tt setShow} function (given in Appendix):
\begin{verbatim}
nat2hfs 42
  S [S [U 0],S [U 0,S [U 0]],S [U 0,S [S [U 0]]]]
hfs2nat (nat2hfs 42)
  42
setShow 42
  "{{{}},{{},{{}}},{{},{{{}}}}}"
\end{verbatim}
\noindent Assuming {\tt urelement\_limit=3} the HFS representation becomes:
\begin{verbatim}
nat2hfs 42
  S [U 0,U 1,U 2,S [U 1]]
setShow 42
  "{0,1,2,{1}}"
\end{verbatim}
\noindent Note that {\tt setShow n} will build a string representation 
of $n \in Nat$, ``shapeshifted" as a $HFS$ 
with Urelements. 
Figure \ref{f42} shows directed acyclic graphs obtained by merging shared 
nodes in the rose tree representation of the $HFS$ associated to 
a natural number (with arrows pointing from sets to their elements).
% HFIGS or VFIGS(Label,Title,Title1,Title2,PNG1,PNG2)
%\VFIGS{f42}{Hereditarily Finite Sets associated to 42}
%{as a pure $HFS$}{with Urelements 0,1,2}{42}{42u3}
\FIG{f42}{Hereditarily Finite Set associated to 42}{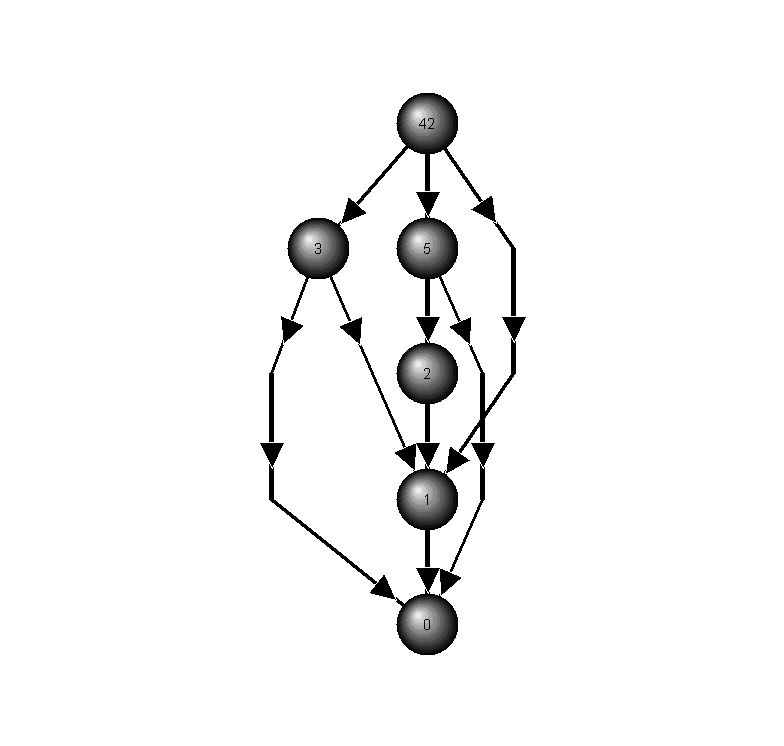}

\subsection{Combinatorial Generation as Iteration}
Using the inverse of Ackermann's encoding, the infinite stream $HFS$ 
can be generated simply by 
iterating over the infinite stream {\tt [0..]}:
\begin{code}
iterative_hfs_generator = map nat2hfs [0..]
\end{code}
\begin{verbatim}
take 5 iterative_hfs_generator
  [U 0,S [U 0],S [S [U 0]],
   S [U 0,S [U 0]],S [S [S [U 0]]]]
\end{verbatim}

\subsection{Generating the Stream of Hereditarily Finite Sets Directly} \label{gen}
To fully appreciate the elegance and simplicity of the combinatorial generation 
mechanism described previously, we will also provide a ``hand-crafted'' recursive 
generator for $HFS$. The reader will notice that this uses some fairly high level 
Haskell constructs like list comprehensions and lazy evaluation, and that in a 
language without such features the algorithm might get significantly more intricate. 

If $P(x)$ denotes the powerset of $x$, the Universe of Hereditarily Finite Sets $HFS$ 
is constructed inductively as follows:
\begin{enumerate}
\item the empty set \{\} is in $HFS$
\item if $x$ is in $HFS$ then the union of its power sets $P^k(x)$ is in $HFS$
\end{enumerate}
\noindent To implement in Haskell a simple $HFS$ generator, conforming this 
definition, we start with a powerset function, working with sets represented as lists:

\begin{code}
list_subsets [] = [[]]
list_subsets (x:xs) = 
  [zs|ys<-list_subsets xs,zs<-[ys,(x:ys)]]
\end{code}  

We can generate the infinite stream of ``pure'' hereditarily finite sets using
Haskell's lazy evaluation mechanism, as follows:
\begin{code}
hfs_generator = uhfs_from 0 where
  uhfs_from k = union (old_hfs k) (uhfs_from (k+1))
  
  old_hfs k = elements_of (hpow k (U 0))
  elements_of (U _) = []
  elements_of (S hs) = hs
 
  hpow 0 h = h
  hpow k h = hpow (k-1) (S (hsubsets h))

  hsubsets (U n) =  []
  hsubsets (S hs) = (map S (list_subsets hs))
\end{code}

One can now extract a finite number of $HFS$ from the stream
\begin{verbatim}
take 5 hfs_generator 
 [S [],S [S []],S [S [S []]],
  S [S [],S [S []]],S [S [S [S []]]]]
\end{verbatim}
and notice the identical behavior of {\tt hfs\_generator} 
and {\tt iterative\_hfs\_generator}. 

\subsection{Encoding Hypergraphs}
\begin{df}
A hypergraph (also called {\em set system}) is a pair $H=(X,E)$ where
$X$ is a set and $E$ is a set of non-empty subsets of $X$.
\end{df}
By limiting recursion to one level in Ackermann's encoding, we can 
derive a bijective encoding of {\em hypergraphs}, 
represented as sets of sets:
\begin{code}
nat2hypergraph = (map nat2set) . nat2set
hypergraph2nat = set2nat . (map set2nat)
\end{code}
as shown in the following example:
\begin{verbatim}
nat2hypergraph 2008
  [[0,1],[2],[1,2],[0,1,2],[3],[0,3],[1,3]]
hypergraph2nat (nat2hypergraph 2008)
  2008
\end{verbatim}
As in the case of $HFS$ combinatorial generation of the infinite 
stream of hypergraphs becomes simply
\begin{verbatim}
map nat2hypergraph [0..]
\end{verbatim}
Note also that a hypothetical application using integers, finite sets and hypergraphs 
can use internally the same immutable data type, with opportunities to {\em
share} common structures.

In the following sections we will think about Ackermann's encoding and its 
inverse as {\em Functors} in Category Theory \cite{PIERCE91}, transporting 
various operations from Natural Numbers to Hereditarily Finite Sets and back.

\section{Shapeshifting Operations between $Nat$ and $HFS$} \label{obv}

\subsection{Fold operators and functors}
Given the {\em rose tree} structure of $HFS$, a natural {\tt fold} 
operation \cite{DBLP:conf/tphol/NipkowP05} can be defined 
on them as a higher order Haskell function:
\begin{code}
hfold f g (U n) = g n
hfold f g (S xs) = f (map (hfold f g) xs)
\end{code}
For instance, it can count how many sets occur in a given $HFS$, as follows:
\begin{code}
hsize = hfold f g where
  f xs = 1+(sum xs)
  g _ =1
\end{code}
Note that recursing over {\tt nat2set} has been used to build a member 
of $HFS$ from a member of $Nat$. Thus, we can combine it with the action 
of a {\tt fold} operator working directly on natural 
numbers as follows:
\begin{code}
nfold f g n = nfold_ f g urelement_limit n

nfold_ f g ulimit n | n<ulimit = g n
nfold_ f g ulimit n = 
  f (map (nfold_ f g ulimit) (nat2set n))
\end{code}
For instance, {\tt nfold} allows counting the elements contained in the $HFS$ 
representation of a number:
\begin{code}
nsize = nfold  f g where 
  f xs = 1+(sum xs)
  g _ =1
\end{code}
as if defined by
\begin{code}
nsize_alt n = hsize (nat2hfs n)
\end{code}

The action of the Ackermann encoding as a Functor from $HFS$ to $Nat$ on morphisms
(seen as functions on a list of arguments) is defined as follows:
\begin{code}
toNat f = nat2hfs . f . (map  hfs2nat)
\end{code}
The same, acting on 1 and 2 argument operations is:
\begin{code}
toNat1 f i = nat2hfs (f (hfs2nat i))
toNat2 f i j = nat2hfs (f (hfs2nat i) (hfs2nat j))
\end{code}
The inverse Ackermann encoding acts as a Functor from $Nat$ to
$HFS$:
\begin{code}
toHFS f = hfs2nat . f . (map nat2hfs)
\end{code}
with variants acting on a 1 and 2 argument functions:
\begin{code}
toHFS1 f x = hfs2nat (f (nat2hfs x))
toHFS2 f x y = hfs2nat (f (nat2hfs x) (nat2hfs x))
\end{code}
Note that the {\tt nat2set} and {\tt set2nat} functions used in the Ackerman 
encoding and its inverse can also be seen as providing Functors 
connecting $Nat$ and $[Nat]$ (seen as a representation of finite subsets of $Nat$):
\begin{code}
toExps f = set2nat . f . (map nat2set)
fromExps f = nat2set . f . (map set2nat)
\end{code}

\subsection{Mappings between Arithmetic and Set Operations} \label{functors}
After extending 2 argument set operations to lists, using {\tt foldl}
\begin{code}
setOp f []=[]
setOp f (x:xs) = foldl f x xs
\end{code}
we can define the equivalent of adduction
(i.e. $\{i\} \cup s$ - see \cite{kaye07,DBLP:journals/mlq/Kirby07}), union,
intersection etc., on natural numbers seen as (lists of) sets:

\begin{code}
nat_adduction i is = 
  set2nat (union [i] (nat2set is))
  
nat_singleton i = 2^i

nat_intersect = nats_intersect . nat2set 
nats_intersect = toExps (setOp intersect)

nat_union = nats_union . nat2set 
nats_union = toExps (setOp union)

nat_equal i j = if i==j then 1 else 0
\end{code}

Similarly, we can transport from $Nat$ to $HFS$, operations like
successor, addition, product, equality as follows:
\begin{code}
hsucc = toNat1 succ
hsum = toNat sum
hproduct = toNat product
hequal = toNat2 nat_equal
hexp2  = toNat1 (2^)
\end{code}
with the practical idea in mind that one can pick the most efficient 
(or the simpler to implement) of the two representations at will.

As current computer architectures tend to support Natural Numbers and 
underlying arbitrary integer representations quite well, we can pick
them as the hub that mediates the ``shapeshiftings'' between 
various data types.
However, in an application where lazy structure building would be 
instrumental for performance, 
something like $HFS$ (or one of the encodings described in 
the next sections) 
could be the most appropriate internal representation. 
 
\section{Pairing Functions} \label{pairings}

{\em Pairings} are bijective functions $Nat \times Nat \rightarrow Nat$.  
Following the classic notation for pairings of \cite{robinson50}, given the 
pairing function $J$, its left and right inverses $K$ and $L$ are such that

\begin{equation}
J(K(z),L(z))=z
\end{equation}

\begin{equation}
K(J(x,y))=x
\end{equation}

\begin{equation} 
L(J(x,y))=y 
\end{equation}

We refer to  \cite{DBLP:journals/tcs/CegielskiR01} for a typical use in the 
foundations of mathematics and to \cite{DBLP:conf/ipps/Rosenberg02a} for an 
extensive study of various pairing functions and their computational properties. 

On top of the ``set operations'' defined in subsection \ref{functors} on $Nat$, 
the classic Kuratowski ordered pair
\begin{math}
(a,b) = \{\{a\},\{a,b\}\}
\end{math}
can be implemented with adductions and singletons as follows:
\begin{code}
nat_kpair x y = nat_adduction sx ssxy where
  sx = nat_singleton x
  sy = nat_singleton y
  sxy = nat_adduction x sy
  ssxy = nat_singleton sxy
\end{code}
However, the Kuratowski pair only provides an injective 
function $Nat \times Nat \rightarrow Nat$, resulting in fast 
growing integers very quickly: 
\begin{verbatim}
[nat_kpair x y|x<-[0..3],y<-[0..3]]
   [2,10,34,514,12,4,68,1028,48,
   80,16,4112,768,1280,4352,256]
\end{verbatim}

\subsection{Cantor's Pairing Function}

We can do better by borrowing some interesting 
pairing functions defined on natural numbers. 
Starting from Cantor's pairing function
bijections from $Nat \times Nat$ to $Nat$ have been used for various proofs 
and constructions of mathematical objects 
\cite{robinson50,robinson55,robinson68a,robinsons68b,DBLP:journals/tcs/CegielskiR01}. 

Cantor's pairing function is defined as:
\begin{code}
nat_cpair x y = (x+y)*(x+y+1) `div` 2+y
\end{code}
Note that its range is more compact
\begin{verbatim}
[nat_cpair i j|i<-[0..3],j<-[0..3]]
  [0,2,5,9,1,4,8,13,3,7,12,18,6,11,17,24]
\end{verbatim}
\noindent Unfortunately, its inverse involves floating point operations
that do not combine well with arbitrary length integers.
\subsection{A new Pairing Function} \label{BitMerge}

We will introduce here a new pairing function, that provides compact
representations for various set theoretic constructs involving ordered pairs
while only using elementary integer arithmetic operations.

Our bijection {\tt bitmerge\_pair} from $Nat \times Nat$ to $Nat$ 
and its inverse {\tt to\_pair} are defined as follows:
\begin{code}
bitmerge_pair (i,j) = 
  set2nat ((evens i) ++ (odds j)) where
    evens x = map (*2) (nat2set x)
    odds y = map succ (evens y)
  
bitmerge_unpair n = (f xs,f ys) where 
  (xs,ys) = partition even (nat2set n)
  f = set2nat . (map (`div` 2))
\end{code}
The function {\tt bitmerge\_pair} works by splitting a number's 
big endian bitstring representation into odd and even bits while 
its inverse {\tt to\_pair} blends the odd and even bits back together. 
With help of the function {\tt to\_rbits} given in Appendix, 
that decomposes $n \in Nat$ into a list of bits (smaller units first) 
on can follow what happens, step by step:
\begin{verbatim}
to_rbits 2008
  [0,0,0,1, 1,0,1,1, 1,1,1]
bitmerge_unpair 2008
  (60,26)
to_rbits 60
  [0,0, 1,1, 1,1]
to_rbits 26
  [0,1, 0,1, 1]  
bitmerge_pair (60,26)
  2008
\end{verbatim}

Note also the significantly more compact packing, compared to Kuratowski 
pairs, and, like Cantor's pairing function, similar growth in both arguments:
\begin{verbatim}
map bitmerge_unpair [0..15]
  [(0,0),(1,0),(0,1),(1,1),(2,0),(3,0),
  (2,1),(3,1),(0,2),(1,2),(0,3),(1,3),
  (2,2),(3,2),(2,3),(3,3)]
[bitmerge_pair (i,j)|i<-[0..3],j<-[0..3]]
  [0,2,8,10,1,3,9,11,4,6,12,14,5,7,13,15]
\end{verbatim}

\subsection{Powersets, Ordinals and Choice Functions} \label{ords}
A concept of (finite) {\em powerset} can be associated to a 
number $n \in Nat$  by computing the powerset of the $HFS$ 
associated to it:

\begin{code}
nat_powset i = set2nat 
  (map set2nat (list_subsets (nat2set i)))
\end{code}
or, directly, as in \cite{abian78}:
\begin{code}
nat_powset_alt i = product 
  (map (\k->1+2^(2^k)) (nat2set i)) 
\end{code}

The von Neumann {\em ordinal} associated to a $HFS$, defined with 
interval notation as $\lambda=[0,\lambda)$, is implemented by the 
function {\tt hfs\_ordinal}, simply by transporting it from $Nat$:
\begin{code}
nat_ordinal 0 = 0
nat_ordinal n = 
  set2nat (map nat_ordinal [0..(n-1)])

hfs_ordinal = nat2hfs . nat_ordinal
\end{code}

The following example shows the {\em transitive structure} of a 
von Neumann ordinal's set representation (see Fig. \ref{f4}). 
It also shows its fast growing $Nat$ encoding ($4 \rightarrow 2059$) 
which can be seen as a somewhat unusual injective embedding of 
finite ordinals in $Nat$, seen as the set of finite cardinals.

\begin{verbatim}
hfs_ordinal 4
  S [S [],S [S []],S [S [],
     S [S []]],S [S [],S [S []],
     S [S [],S [S []]]]]
nat_ordinal 4
  2059
\end{verbatim}

\VFIGS{f4}{4 and its associated ordinal}{as a pure $HFS$}{its associated ordinal 2059}{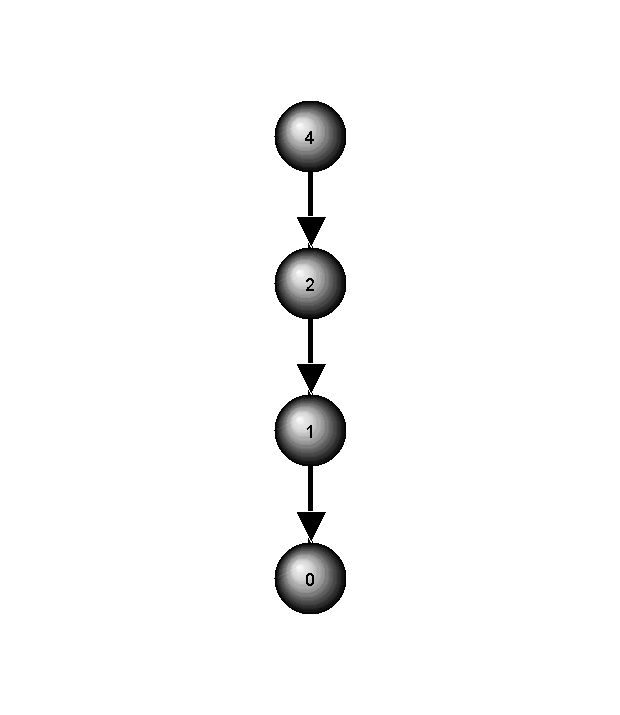}{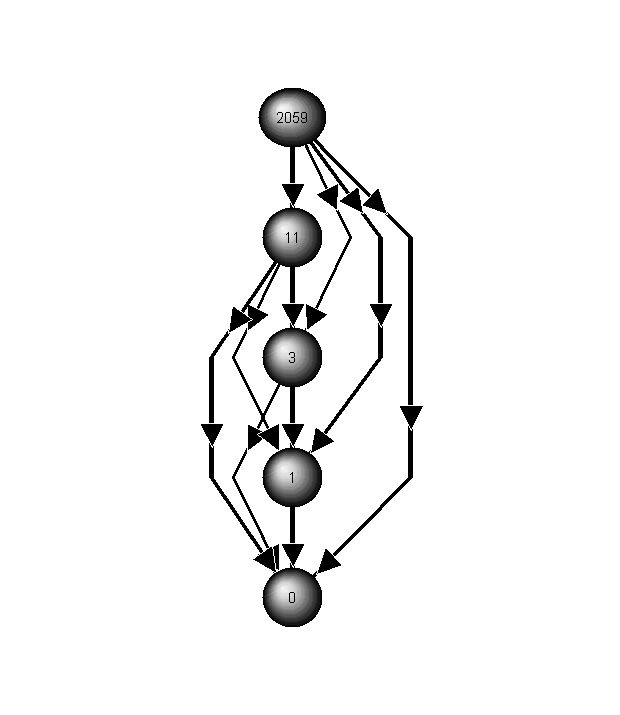}

Finally, a choice function
is implemented as an encoding of pairs of sets and their first 
elements with our compact $Nat \times Nat \rightarrow Nat$ 
pairing function:
\begin{code}
nat_choice_fun i =  set2nat xs where 
  es = nat2set i
  hs = map (head . nat2set) es
  xs = zipWith (curry bitmerge_pair) es hs
\end{code}
As {\em even} numbers represent sets that do not contain the 
empty set as an element, we  compute $Nat$ representations 
of the choice function as follows:
\begin{verbatim}
map nat_choice_fun [0,2..16]
  [0,2,64,66,32,34,96,98,16777216]
\end{verbatim}
Note that {\tt nat\_choice\_function} computes a natural number representation
{i.e. G\"{o}edel number} for a function that picks an element of each set of
any family of sets not containing the empty set. 
Constructing such a natural number proves that $Nat$, with the
structure borrowed from $HFS$ is actually {\em a model}
for the {\em Axiom of Choice}. Such models are important
in the foundations of mathematics as they show that
interpretations of sets and functions other the usual ones
are compatible with various axiomatizations 
of set theory \cite{kaye07,DBLP:journals/mlq/Kirby07}.

\section{Directed Graph Encodings} \label{graphs}

Directed Graphs are equivalent to binary relations seen as 
sets of ordered pairs. Equivalently, (as implemented in the 
Haskell {\tt Data.Graph} package), they can also be seen as 
arrays of vertices in [0..n] paired with lists of vertices 
of adjacent outgoing edges. We will freely alternate between 
these two representations in this section.

\subsection{Directed Acyclic Graph representations for $HFS$} \label{graphrep}

The {\em rose tree} representation of $HFS$ can be seen as a {\em set} 
of edges, oriented to describe either set membership $\in$ 
or its transpose, set containment.
\begin{code}
nat2memb = nat2pairs with_memb
nat2contains = nat2pairs with_contains

with_memb a x = (x,a)
with_contains a x = (a,x)
\end{code}
Note that this uses the function {\tt nat2pairs} (see Appendix) that 
provides the actual decomposition of a number into Haskell ordered pairs. 
The following examples show how this works:
\begin{verbatim}
nat2memb 42
  [(0,1),(0,3),(0,5),(1,2),(1,3),
   (1,42),(2,5),(3,42),(5,42)]
nat2contains 42
  [(1,0),(2,1),(3,0),(3,1),(5,0),
   (5,2),(42,1),(42,3),(42,5)]
\end{verbatim}
These list of pair representations can be easily converted to
Haskell's graph data type (imported from Data.Graph) as follows:
\begin{code}
nat2member_dag = nat2dag_ nat2memb 

nat2contains_dag = nat2dag_ nat2contains   

nat2dag_ f n = buildG (0,l) es where 
  es=reverse (f n)
  l=foldl max 0 (nat2parts n)
\end{code}
where {\tt nat2parts}, given in the Appendix, converts $n$ to the set of 
Natural Numbers occurring in its $HFS$ representation.

Moreover, the pair representation of $\in$ and its inverse can be turned into
a more compact graph by replacing its $n$ distinct vertex 
numbers with smaller integers 
from $[0..n-1]$, by progressively building a map describing this 
association, as shown in the function {\tt to\_dag}
\begin{code}
to_dag n = (buildG (0,l) 
  (map (remap m) (nat2contains n))) where  
    is = [0..l]
    ns =  reverse (nat2parts n)
    l = (genericLength ns)-1
    m= (zip ns is)
    remap m (f,t) =  (lf,lt) where 
      (Just lf)=(lookup f m) 
      (Just lt)=(lookup t m)
\end{code}

Dually, one can convert $n \in Nat$ to the containment graph 
of its $HFS$ as follows
\begin{code}
to_ddag = transposeG . to_dag
\end{code}

An interesting question arises at this point. 
{\em Can we rebuild a natural number from its directed acyclic
graph representation, assuming no labels are available, except 0?} 
Surprisingly, the answer is yes, and the function {\tt from\_dag} 
provides the conversion:
\begin{code}   
from_dag g = 
  compute_decoration g (fst (bounds g))

compute_decoration g v = 
  compute_decorations g (g!v) where
    compute_decorations _ [] = 0
    compute_decorations g es =
      sum (map ((2^) . (compute_decoration g)) es)
\end{code}
\begin{verbatim}
to_dag 42
  array (0,5) [(0,[1,2,4]),(1,[3,5]),(2,[4,5]),
               (3,[4]),(4,[5]),(5,[])]
from_dag (to_dag 42)
  42
to_ddag 42
  array (0,5) [(0,[]),(1,[0]),(2,[0]),(3,[1]),
               (4,[3,2,0]),(5,[4,2,1])]
\end{verbatim}
After implementing this function, we have found that it closely 
follows the {\em decoration} functions used in Aczel's book \cite{aczel88}, 
and renamed it {\tt compute\_decoration}. In the simpler case of the $HFS$ universe, 
with our well-founded sets represented as DAGs, the existence and unicity of the 
result computed by {\tt from\_dag} follows immediately from the 
Mostowski Collapsing Lemma (\cite{aczel88}).

\subsection{Extensional/Intensional Duality} \label{dual}

What can be said about the graphs obtained by reversing 
the direction of the arrows representing the $\in$ relation? 
Intuitively, it corresponds
to the fact that intensions/concepts 
would become the building blocks of the theory, provided 
that something similar to the {\em axiom of extensionality} holds. 
In comments related to Russell's type theory \cite{russelcrit} 
pp. 457-458 G\"{o}del mentions an {\em axiom of intensionality} with 
the intuitive meaning that ``different definitions belong to
different notions". G\"{o}del also notices the duality between 
``no two different properties belong to exactly the same things" 
and ``no two different things have exactly the same properties"  
but warns that contradictions in a simple type theory would result 
if such an axiom is used non-constructively.

We can now look for the presence of intensional/extensional symmetry 
in $HFS$ by trying to rebuild a $HFS$ 
representation from the transpose of $\in$, $\ni$:
\begin{code}
from_ddag g = 
  compute_decoration g (snd (bounds g))

intensional_dual_of = from_ddag . to_ddag
\end{code}
Are such representations self-dual? Let's define as {\em self-dual} 
a number $n \in Nat$ that equals its intensional dual and then 
filter self-dual numbers in an interval:
\begin{code}
self_idual n = n==intensional_dual_of n

self_iduals from to = 
  filter self_idual [from..to]  
\end{code}
Unfortunately, as the following example shows, relatively 
few numbers are self-duals:
\begin{verbatim}
self_iduals 0 1000
  [0,1,2,3,4,5,10,11,16,17,34,35,
   64,65,130,131,264,265,522,523]
\end{verbatim}
Figures \ref{iduals1} and \ref{iduals2} 
show some $HFS$ graphs of natural numbers equal to their 
intensional duals.
\VFIGS{iduals1}{self-dual and its intensional dual 
as $HFS$ graphs}{131}{the dual of 131}{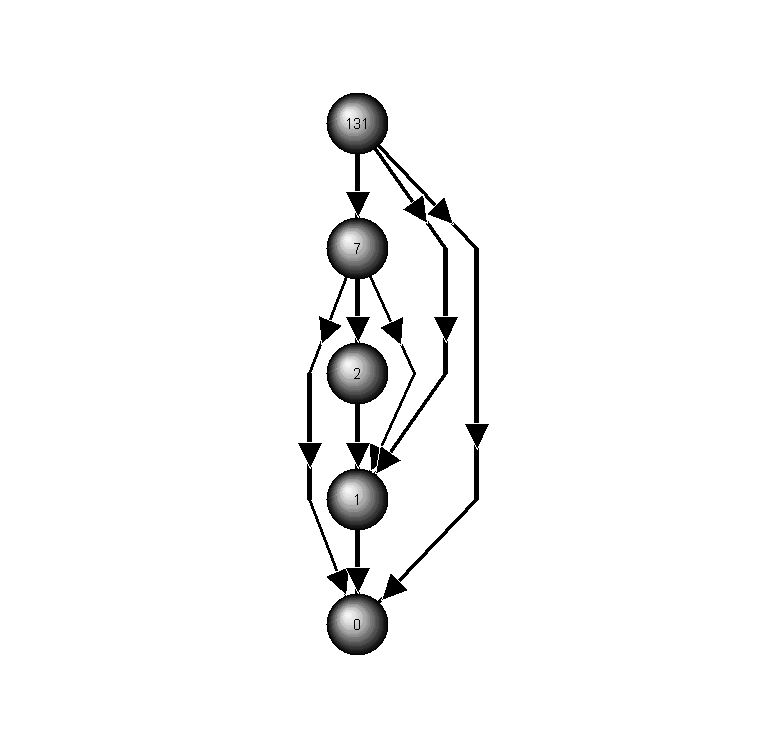}{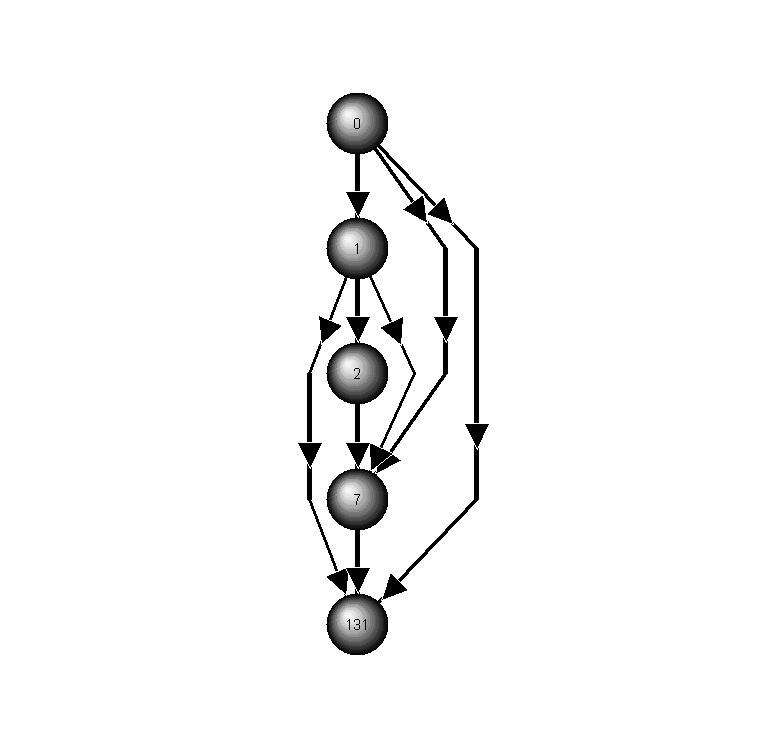}
\VFIGS{iduals2}{self-dual and its intensional dual 
as $HFS$ graphs}{16393}{the dual of 16393}{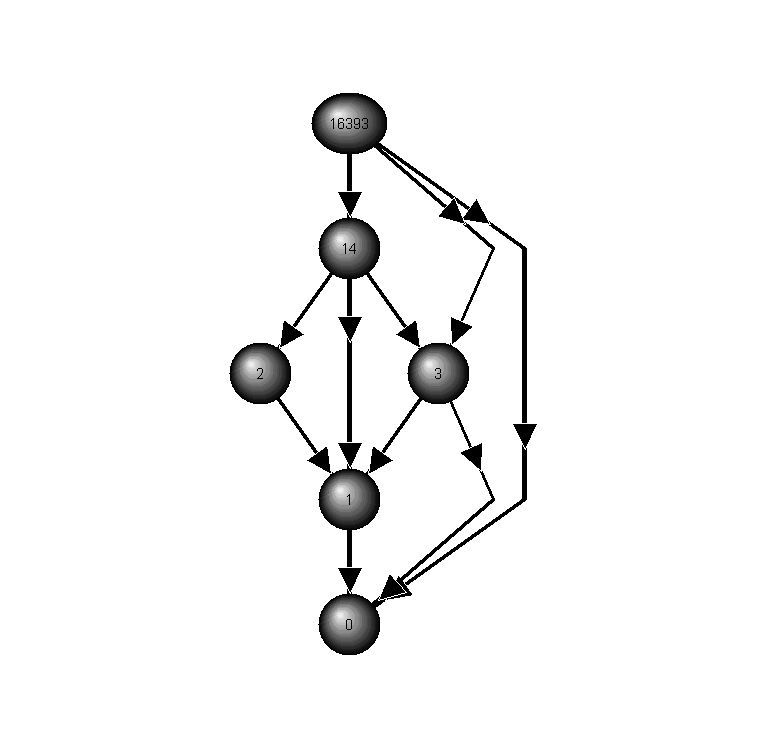}{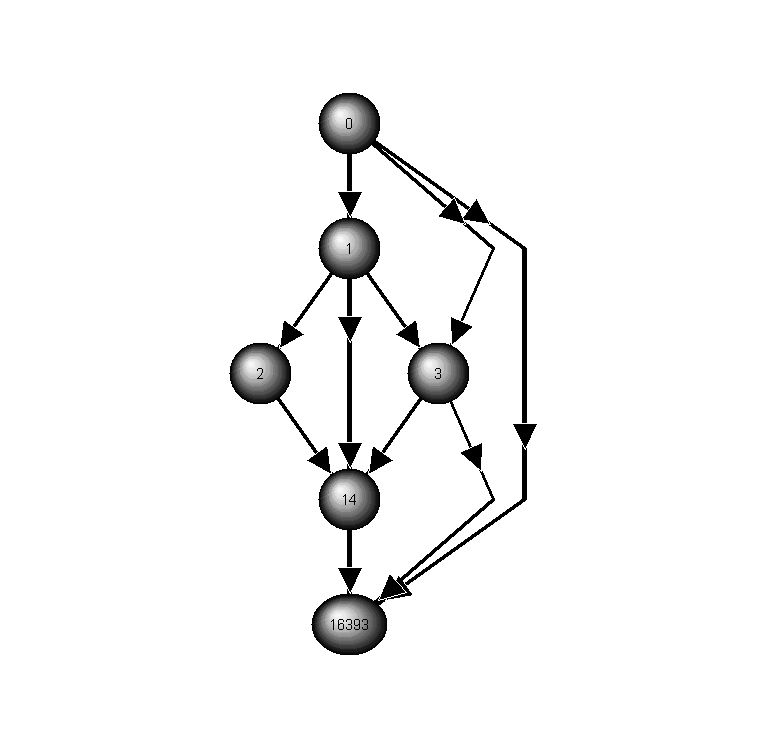}

We will leave it
as a topic for future research to investigate more in depth,
various aspects of $\in$ / $\ni$ duality in $HFS$, in correlation 
with Natural Numbers and their encodings.

\subsection{Encodings of Directed Graphs as Natural Numbers} \label{digraphs}

Hypersets \cite{aczel88} are defined by replacing the 
Foundation Axiom with the AntiFoundation axiom. 
Intuitively this means that the $\in$-graphs can 
be cyclical \cite{barwise96}, provided that 
they are minimized through 
{\em bisimulation equivalence} \cite{DBLP:conf/cav/DovierPP01}.
We have not (yet) found an elegant encoding of hereditarily finite 
hypersets as natural numbers, similar to Ackerman's encoding. 
The main difficulty seems related to the fact that hypersets are 
modeled in $HFS$ as equivalence classes with respect to 
bisimulation \cite{aczel88,barwise96,DBLP:journals/tplp/PiazzaP04}. 
Toward this end, an easy first step seems to find a bijection from 
directed graphs (with no isolated vertices, corresponding to their
 view as binary relations), to $Nat$:
\begin{code}
nat2digraph n = map bitmerge_unpair (nat2set n)
digraph2nat ps = set2nat (map bitmerge_pair ps)
\end{code}
With digraphs represented as lists of edges, this bijection 
works as follows:
\begin{verbatim}
nat2digraph 2008
  [(1,1),(2,0),(2,1),(3,1),
   (0,2),(1,2),(0,3)]
digraph2nat (nat2digraph 2008)
  2008
nat2digraph (255)
  [(0,0),(1,0),(0,1),(1,1),
  (2,0),(3,0),(2,1),(3,1)] 
digraph2nat (nat2digraph 255)
  255  
\end{verbatim}
As usual
\begin{verbatim}
map nat2digraph [0..]
\end{verbatim}
provides a combinatorial generator for the infinite stream of 
directed acyclic graphs.

\section{Related work} \label{related}
Natural Number encodings of Hereditarily Finite Sets have triggered 
the interest of researchers in fields ranging from Axiomatic Set Theory 
and Foundations of Logic to Complexity Theory and Combinatorics
\cite{finitemath,kaye07,DBLP:journals/mlq/Kirby07,abian78,DBLP:journals/mlq/Kirby07,DBLP:journals/jct/MeirMM83,DBLP:conf/foiks/LeontjevS00,DBLP:journals/tcs/Sazonov93,avigad97}. 
Graph representations of sets and hypersets based on the variants of the 
Anti Foundation Axiom have been studied extensively in \cite{aczel88,barwise96}.
Computational and Data Representation aspects of Finite Set Theory and 
hypersets have been described in logic programming and theorem proving contexts in 
\cite{dovier00comparing,DBLP:journals/tplp/PiazzaP04,DBLP:conf/cav/DovierPP01,DBLP:conf/types/Paulson94}. 
Pairing functions have been used work on decision problems as early as
\cite{pepis,kalmar1,robinson50,robinson55,robinson68a,robinsons68b}. 
Various mappings from natural number encodings to Rational Numbers are described 
in \cite{rationals}, also in a functional programming framework.

\section{Conclusion and Future Work} \label{concl}

Implementing with relative ease the encoding techniques
typically used only in the foundations of mathematics recommends Haskell
as a surprisingly effective tool for experimental mathematics.

We have described a variety 
of isomorphisms between mathematically interesting data structures, all centered 
around encodings as Natural Numbers. The possibility of sharing significant
common parts of HFS-represented integers could be used in implementing shared 
stores for arbitrary length integers. Along the same lines, another application 
would be data compression using 
some ``information theoretically minimal" variants of the graphs in 
subsection \ref{graphrep}, from which larger, $HFS$ and/or 
natural numbers can be rebuilt.

Last but not least, making more accessible to computer science 
students some of the encoding techniques typically used only 
in the foundations of mathematics (and related reasoning techniques), 
suggests applications to teaching discrete mathematics and/or 
functional languages in the tradition of \cite{Hall00}.

\bibliographystyle{plainnat}
\bibliography{INCLUDES/theory,tarau,INCLUDES/proglang,INCLUDES/biblio,INCLUDES/syn}

%\appendix
\section*{Appendix}
To make the code in the paper fully self contained, 
we list here some auxiliary functions.

\paragraph{Bit crunching functions} 
The following functions implement conversion operations between bitlists and numbers.
Note that our bitlists represent binary numbers by selecting exponents of 2 in 
increasing order (i.e. ``right to left"). 
\begin{code}
-- from decimals to binary as list of bits
to_rbits n = to_base 2 n

-- from bits to decimals
from_rbits bs = from_base 2 bs

-- conversion to base n, as list of digits
to_base base n = d : 
  (if q==0 then [] else (to_base base q)) where
    (q,d) = quotRem n base

-- conversion from any base to decimal 
from_base base [] = 0
from_base base (x:xs) = x+base*(from_base base xs)
\end{code}

\paragraph{String Representations}
The function {\tt setShow} provides a string representation 
of a natural number as a ``pure" HFS.
\begin{code}
setShow n = sShow urelement_limit n
\end{code}
The function {\tt sShow} provides a string representation of 
a natural number as a HFS with {\em Urelements}.
\begin{code}
sShow 1 0 = "{}"
sShow ulimit n | n<ulimit = show n
sShow ulimit n = "{"++ 
  foldl (++) "" 
    (intersperse "," (map (sShow ulimit) (nat2set (n-ulimit)))) 
  ++"}"
\end{code}

\paragraph{Conversion to Ordered Pairs}
The function {\tt nat2pairs} converts a natural number to a 
set of Haskell ordered pairs expressing the $\in$ relation 
on its associated $HFS$ or its dual $\ni$.
\begin{code}
nat2pairs withF n  = (sort . nub)  (nat2ps withF n)

nat2ps withF 0 = []
nat2ps withF from = 
  ((n2rel ns) ++ (ns2rel ns)) where
    f = withF from
    n2rel = map f
    ns2rel = concatMap (nat2ps withF)
    ns=nat2set from 
\end{code}
The function {\tt nat2parts} converts $n$ to the set of Natural Numbers 
occurring in the $HFS$ representation of $n$.
\begin{code}
nat2parts = sort . nub . nat2repeated where
  nat2repeated 0 = [0]
  nat2repeated from = from : (nat2more ns) where
    nat2more = concatMap nat2repeated
    ns=nat2set from 
\end{code}

\begin{comment}
\begin{code}  
mprint f = (mapM_ print) . (map f)
\end{code}
\end{comment}

\end{document}